\documentclass[runningheads]{llncs}

\usepackage{times}
\usepackage{tabularx}
\usepackage{graphicx}
\usepackage{url}
\usepackage[strings]{underscore}
\usepackage{wrapfig}
\usepackage{verbatim}

\newtheorem{defn}{Definition}

\usepackage{listings}
\usepackage[T1]{fontenc}
\usepackage{textcomp}
\usepackage[variablett]{lmodern}

\begin{document}

\title{Solving Financial Regulatory Compliance\\ Using Software Contracts}
 
\author{Newres Al Haider \and
  Dilhan Thilakarathne \and
  Joost Bosman }

\authorrunning{N. Al Haider et al.}

\institute{ING Tech Research and Development, \\
ING Bank, Haarlerbergweg 13-19. 1101 CH Amsterdam The Netherlands
  \email{\{Newres.Al.Haider,Dilhan.Thilakarathne,Joost.Bosman\}@ing.com}}

\maketitle

\begin{abstract}

Ensuring compliance with various laws and regulations is of utmost priority for
financial institutions. Traditional methods in this area have been shown to be
inefficient. Manual processing does not scale well. Automated efforts are
hindered due to the lack of formalization of domain knowledge and problems of
integrating such knowledge into software systems.  In this work we propose an approach to tackle these issues by encoding them into software contracts using a Controlled Natural Language. In particular, we encode a portion of the Money Market Statistical Reporting (MMSR) regulations into contracts specified by the clojure.spec framework. We show how various features of a contract framework, in particular clojure.spec, can help to tackle issues that occur when dealing with compliance: validation with explanations and test data generation. We benchmark our proposed solution and show that this approach can effectively solve compliance issues in this particular use case.


\end{abstract}

\section{Introduction \& Motivation}\label{sec:introduction}

In a sustainable society, financial institutions cannot neglect their
responsibilities of greater accountability, adequate transparency and social
trust. Financial institutes may differ in the way they operate, but all of them
must comply with regulations to fulfill the above criteria. In addition to
non-compliance affecting the image of the institution, it may also result in
judiciary prosecution with significant fines. As per the Boston Consulting Group
(BCG) report~\cite{grasshoff2017global}, the total cost in penalties
through regulatory enforcement on banks, during the period of 2009 through 2016, is roughly
\$321 billion worldwide.

In response to these challenges financial institutes are expanding their
compliance departments with the necessary experts to ensure that they are fully
compliant. However, due to the rapidly increasing volumes of new regulations, as
well as the amount of data they pertain to, ensuring compliance remains an increasingly costly issue in both time and expertise.

Automating the compliance process is a difficult task that requires solving a
large number of challenges. Domain expertise has to be used to translate legal and
regulatory texts into a more formalized form that can enable automation. This model
then would need to be embedded into a software system that is capable of
reliably and effectively enable compliance checking. In addition this has to be
done in a way that is understandable to domain experts as well as to auditors.

The compliance process itself does not stand alone from the rest of
the financial institution. To ensure compliance at all levels, compliance to
regulations has to be an integrated element of all products and processes. In a
modern bank software systems are an integral part of providing financial
services. This means that the software itself has to be designed and maintained in accordance with compliance regulations. 

Traditional software engineering techniques and methods often fail with regards
to integrating compliance requirements into the full software development
life-cycle. It has been noted that compliance is often tacked on in a `Big Bang'
manner, just before the release of the overall product, as opposed to the more
desired `continuous compliance' approach~\cite{fitzgerald2014continuous}.

While various model-based approaches to compliance have been proposed, they are
often hampered in real-life applicability due to a number of reasons. A notable
one of these is the lack of the ability to incorporate arbitrary complex compliance patterns~\cite{becker2012generalizability}.

Another problem for testing automated compliance solutions is that real-world
data, on which the automated compliance process would be tested, is often difficult
to obtain. Such data can also be the subject of compliance regulations, such as
the General Data Protection Regulation (GDPR)~\cite{eu-679-2016} in the EU.

One common element in all of these issues, is that it denotes a difficulty
between the various stakeholders and domains of expertise to communicate what is required to effectively automate compliance regulations.
 
In this work, we propose an approach that aims to alleviate these problems by
encoding the compliance regulations into contract
systems~\cite{meyer1992applying}. These systems were originally designed to
guide software development by encoding the correct program behavior. This makes
them a suitable tool to enable `continuous compliance' in software systems. In
addition, some contract systems can use arbitrary functions, that return a
boolean value, to define the correct behavior. This allows for the
incorporation for very complex compliance patterns. Such systems
can also provide the necessary features to give detailed explanations why a
particular regulation did, or did not, hold. Finally, there exist
contract systems that have a tight integration with generative test frameworks,
notably the `clojure.spec'~\cite{clojure-spec} library. Such tools can be adapted to generate relevant test data for compliance regulations and thereby
raising the confidence in their implementation.

To make this encoding process transparent to domain experts as well as software engineers alike, we make use of a Controlled Natural Language (CNL). CNLs are subsets of natural
languages that have been designed to reduce or eliminate ambiguities that could
occur when interpreting natural language texts. In our approach they provide a
common language as an intermediate step between the natural language of
regulations and the executable software contracts. This is done in order to
ensure that relevant portions of the regulations are made understandable for all
parties involved. Controlled Natural Languages have been
used previously to make formal tools and methods more accessible, such as
Attempto Controlled English for semantic web based knowledge representations~\cite{attempto1}, or more recently for Answer Sets \cite{fang2017approach}.

To demonstrate this approach, we implement a portion of the Money
Market Statistical Reporting (MMSR) regulations
\cite{eu-1333-2014,mmsr-reporting} using the `clojure.spec' \cite{clojure-spec}
library in the Clojure programming language \cite{hickey2008clojure}. We show
that regulations can be implemented and checked using software contracts, namely
`clojure.spec', for this use case. Valid examples of messages also can be
generated by using the generator framework associated with `clojure.spec'. This
process is benchmarked in conjunction with validation of such generated examples.

The rest of this paper is structured as follows: first, in
Section~\ref{sec:softwarecontracts} we introduce some background on software
contracts, and in particular clojure.spec. In Section~\ref{sec:approach} we
describe our approach in more detail. In Section~\ref{sec:usecase} we describe
how we encoded a portion of the MMSR regulations using our approach into contracts and how we used
this to perform our evaluation. Related work and how they compare to our
approach can be found in Section~\ref{sec:related}. Finally, we conclude with Section~\ref{sec:conclusion} where we summarize our findings and outline future research.

\section{Software Contracts}\label{sec:softwarecontracts}

Making various assertions about the function of a computer program and using
these to facilitate the correctness of it has a long history
\cite{hoare1969axiomatic}. Basic examples of such statements are 
things that must be true before or after the execution of a program,
preconditions and postconditions, respectively.

The explicit link with the use of such assertions and the notion of contracts
was made by Meyer \cite{meyer1992applying}. In this work it is detailed that,
similarly to traditional legal contracts, assertions can be used to specify the
obligations and benefits that a particular (piece of) program can provide.
These specifications can be invaluable when designing correct programs. They can
also be used as a tool to document and communicate the specifications of software in an
explicit manner~\cite{jazequel1997design}. Systems built around the use of such
contracts, i.e. contract systems, are available as part of a number of
programming languages, such as Eiffel \cite{meyer1992eiffel}, D
\cite{alexandrescu2010d}, Racket \cite{felleisen2015racket} and Clojure
\cite{hickey2008clojure}. In other languages such facilities can be provided
using an external library, for example in Java \cite{leavens2006preliminary} and
in .NET languages~\cite{fahndrich2010static}.

In contract systems, the `contract' itself is known by various terminology, such
as: `contract' \cite{meyer1992eiffel,felleisen2015racket}, `schema'
\cite{plumatic-schema} and `spec' \cite{clojure-spec}. In this work, unless
explicitly referring to a specific type of contract system, we use the term
`software contract' or simply `contract' to refer to such contracts in general, and `contract system' or `software contract systems' to the frameworks using them.

While there are many types of contract systems, in this work, we will make use
of the `clojure.spec' library which is part the Clojure language. Clojure is a
Lisp as well as a dynamic, functional, data-oriented language
\cite{hickey2008clojure}. In Clojure, information is represented as (immutable)
data directly in the form of various `primitive' values: numbers, strings, keywords and collections such as
vectors, lists, sets, maps of these values (or other collections). This is in
contrast with some other languages that encapsulate such information into various abstractions such as objects.

However, especially in the light that it is a dynamically typed language, it can
be of vital importance to communicate the various characteristics of data used as input (preconditions) and output
(postconditions) to a function. The library of `clojure.spec'
\cite{clojure-spec} was introduced as part of the language to help with these
issues. Here we only give a very brief description of the framework, alongside
with some examples in Table~\ref{table:spec}. For more detailed information, we
refer the reader to a more comprehensive guide on this subject \cite{clojure-spec-guide}.
 
\begin{table}
\footnotesize
\begin{center}
    \begin{tabular}{| l |  p{7cm} | p{2cm} |}
    \hline
      Functionality & Example(s) & Evaluates To  \\
\hline
      Registering & (s/def ::fruit \#\{``apple'' ``pear'' ``cherry''\})
       \newline (s/def ::veg \#\{``carrot''  ``cucumber''\})
                              & ::fruit
                                \newline ::veg \\
      Validating  & (s/valid? ::fruit ``apple'') & true \\
      Composing  & (s/def ::fruit-or-veg (s/or :fruit ::fruit :veg ::veg)) & ::fruit-or-veg \\
      Conforming  & (s/conform ::fruit-or-veg ``carrot'') & [:veg ``carrot''] \\
      Generating & (gen/generate (s/gen ::veg))& ``cucumber'' \\
\hline
    \end{tabular}
\end{center}

\caption{Spec use examples, where the namespace `s' refers to `clojure.spec',
  `gen' to the `clojure.spec.generator' namespaces, and any other namespaced keywords to an example namespace.}
  \label{table:spec}
\end{table}

Specs are in essence single parameter functions that evaluate to either true or false. For
example the function `even?' that checks whether a number is even, is itself a
spec. This makes this particular contract system very expressive, and allows for
existing code written for the purpose of validating data to be easily pluggable inside the spec framework.

There are various ways a spec can be utilized. Validation checks whether a given
value fulfills a spec. Conform and explanation allow the system to tell why a particular value conforms or explain why it does not.

Specs themselves can be combined through a number of combinators into other
specs. There are combinators for logical predicates, such as `or' and `and',
regular expression based combinators for describing order such as `*',
`?', `+'. In addition, there are various combinators for expressing
restrictions on collections such as maps, vectors, sets, etc.

Specs can be registered into a registry by using namespaced keywords.
This allows to uniquely refer to specs, similarly how resources are referred to
in RDF~\cite{world2014rdf}. Keywords play a large role in conformance and
explanations, as they allow to refer to a specific part of a composed
contract.

One other feature of this system that we utilize is the generation
library: `clojure.spec.generator'. This allows the user to generate a random,
correct value for a contract. In many straightforward cases the system is able
to handle such generation with no additional input. For more complex specs custom
generators need to be built using this generation library.

Finally, due to space reasons when we refer to the spec or the
generation libraries, we use the namespaces `s' and `gen' respectively as
shorthands, such as in the examples described in Table~\ref{table:spec}.

\section{Approach}\label{sec:approach}

As mentioned in Section~\ref{sec:introduction}, ensuring the automated
compliance process at financial institutions requires expertise from many
domains. Knowledge from the financial, legal and software engineering fields
have to be marshaled in order to have an effective implementation. Various
issues could hinder this automation. These include: being unable to express
arbitrary compliance rules, limited integration of compliance within the
software engineering process and a lack of effective communication between
experts of the different domains. In this section we propose our approach for
using software contracts, and in particular `clojure.spec', which we described
in Section~\ref{sec:softwarecontracts}, to help deal with these issues.

\begin{figure}
  \begin{center}
    \includegraphics[width=1\textwidth]{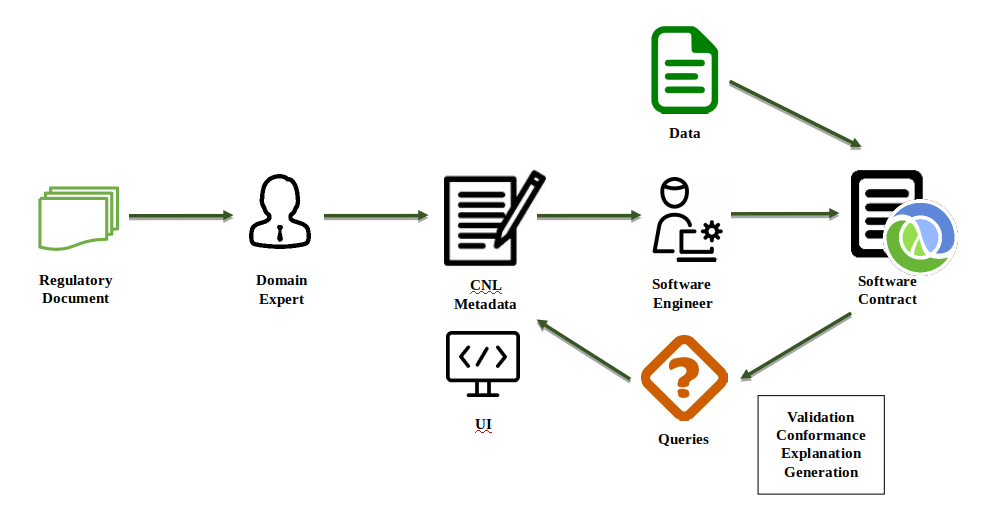}
  \end{center}
  \caption{Regulatory compliance checking through software contracts}
  \label{fig:approachFig}
\end{figure}

An overview of our approach can be seen in Figure~\ref{fig:approachFig}, which
highlights the process and communication lines. The approach is an iterative
process, with the overall aim of building up a software contract that can fully
express and use the intended regulation. An additional aim is to do this in a
manner that makes the automated regulation checking more transparent and
verifiable for domain experts.

A step-by-step description of the approach is as follows:

The process starts with a regulatory document (or documents) containing the
regulations that are needed to be implemented.

First, domain expert(s) identify parts of the document that needs to be
formalized. This selection is made explicitly, as not all text in the document
might be relevant for the problem at hand. In addition, as the process is
iterative, a selection will have to be made on what should be tackled in the
first, or subsequent, step.

The selected text is then formalized with a Controlled Natural Language (CNL). A
CNL-based representation was chosen due to the fact that the CNL is a subset of
the (English) natural language, but without the ambiguities that might occur in
a full natural text. As such it is accessible to legal/financial experts and
software engineers alike. This CNL forms the common language over which domain
experts can communicate with software engineers, about what needs to be
implemented, in order to comply with the regulation.

The CNL is designed to be a formal abstraction of the regulations in a way that
easily maps to the software contracts. It also aims to remain understandable and
writable by domain experts without software engineering expertise. We will
discuss the CNL and the design decisions behind it in the later part of this
section.

A user interface (UI) is integrated into this process to aid the domain experts
in viewing and writing this CNL-based representation. The natural language text
from which the CNL is formalized is added as metadata to this representation.
This ensures that the implementation is fully traceable back to the source text.
This metadata is used as additional information for the full implementation by
the software engineer.

Finally the CNL, along with the metadata, would be translated into the software
contracts (`clojure.spec' specs given our implementation in this paper). In this
process the software engineer would implement the required elements that the
CNL-based abstraction did not yet cover. If for any reason the structure defined
by the CNL would need to change, based on the insight of the software engineer,
changes can be made which would also reflect back in the CNL. The end result
would be an executable version of the compliance regulation in the form of
software contracts.

These contracts are capable of validating regulations, given the relevant data,
while providing explanations. They can also provide features such as data
generation to help with validation of the implementation, as mentioned in
Section~\ref{sec:softwarecontracts}.

The results of querying the contracts and their features are tied back to the
CNL-based representation using the UI. For example, if a particular run has
failed according to the contract, the exact cause can be pinpointed to the
related abstractions in the CNL. This allows the domain expert to have a clearer
view on what is happening with the implementation at the level of abstraction
they previously used for formalizing the regulations. Other features of the
software contract can be tied to this CNL representation as well. Any changes
made to the abstracted portion of the software contracts can be reflected back
to it. Generated data from the contract framework can provide the domain experts
with additional ways of validating the implementation.

This process is iterative. The implemented contracts can be tested by using
actual, or generated, data and by reflecting on whether the resulting CNL
effectively matches the intended regulations and its implementation. Based on
this information changes can be made until all parties are satisfied that the
implementation is a sufficient enough representation of regulations outlined in
the regulatory document.

A key component in the overall process, aside from the software contracts
already explained in Section~\ref{sec:softwarecontracts}, is the CNL which we
will introduce in the remaining part of this section.

\subsection{Controlled Natural Language}\label{sec:cnl}

Here we aim to describe in detail the Controlled Natural Language (CNL) based
representation that we use in our approach. First we describe our design
principles for the CNL. Next we will give a formalisation of the semantics
behind the CNL. Finally we will show how this formalisation maps to both the
actual natural language sentences that make up the CNL as well as to the
software contracts that it abstracts.

The overall purpose of CNL is to be a formal intermediary artifact between the
natural language text of the regulations and the executable artifacts of the
software contracts. When designing the CNL we had the following principles in
mind:

\begin{itemize}
\item The CNL is intended to be an abstraction of the executable software
  contracts, not a complete representation of them. This property was given to
  the CNL as we are in a situation where we have two, somewhat conflicting,
  goals. On one hand we want to express arbitrary compliance rules, as mentioned
  in Section~\ref{sec:introduction}. On the other hand we want to ensure that
  the language can be understood and written by domain experts without software
  engineering knowledge. While making a general-purpose (Turing complete)
  programming language more accessible to domain experts is an interesting
  research problem in itself, we instead have chosen to make the CNL an
  abstraction of the software contracts. This allows for the domain experts to
  work on a level of detail that they are more comfortable with. However this
  means that our model has to abstract away some elements of the full software
  contracts.
\item The CNL is intended to be a sound representation of the software contract
  implementation of the regulation. Although we intend to abstract away some of
  the semantics of the full implementation, we want to ensure that all the
  semantics that are specified with the CNL are fully captured within the
  contracts.
\item Every contract represented by the CNL should have a unique name under
  which it is registered and which should match between the CNL and the actual
  implementation. As mentioned in Section~\ref{sec:softwarecontracts}, in the
  clojure.spec library contracts can be registered under namespaced keywords to
  uniquely refer to them. For the CNL we mandate such a feature. This means
  there should be a unique namespaced name for each software contract abstracted
  by the CNL. For the purpose of this paper, these names are all namespaced
  keywords, as we use the `clojure.spec' library for our implementation. In a
  different contract system other naming structures can be used. This ensures
  that there is always a straightforward mapping between the implementation and
  its abstraction. This mapping can be used when providing feedback from the
  executed system, notably with the explanations.
\item The CNL aims to emulate the tree structure that would underlay the actual
  software contract implementation representing a regulatory document. As
  mentioned in Section~\ref{sec:softwarecontracts}, contracts can be composed
  from other contracts to build up more complex restrictions. At the root level
  there always exists a contract representing all the requirements described by
  the whole regulatory document. In most non-trivial cases this contract would
  be composed of other contracts that are either compound (i.e.: composed of
  other contracts themselves), or atomic (i.e.: no other contract was used in
  defining them). This composition roughly corresponds to the root, inner nodes
  and leaves of a tree. Our CNL aims to abstract away the implementation details
  of the contracts, while keeping the links between the contracts intact. The
  goal of this is to allow the domain expert to outline and view important
  structural elements of the implementation, without having to fully formalize,
  or understand, all the implementation details required.
\end{itemize}

Based on these principles we can give the formal definition of the semantics
behind the CNL as follows:

\begin{defn}[Atomic Element]
  An Atomic Element is a representation of a software contract that we do not
  want to break down into further components within the CNL. Given a software
  contract with a unique name $k$ an Atomic Element is represented by
  $k_{\mathit{atom}}$.
\end{defn}

In some cases these atomic elements are representations of software contracts
that are not combined from others (see Section~\ref{sec:softwarecontracts} for
some examples of combined and not combined contracts). In other cases they can
be software contracts for which we abstract away their internal structure. This
is due to the fact that their implementation details are not relevant, or easily
understandable, from the perspective of the domain experts. A good example of
this would be a requirement from a regulation stating that a date is only valid
if it is given in an ISO 8601 compliant format. We could represent this as the
atomic element $\mathit{validISOdate}_{\mathit{atom}}$ where `validISOdate' is a
unique name (w.r.t our represented regulations). The existence of this
requirement should be reflected within the CNL, but it is not required for the
domain expert to read, and especially write, all the implementation details
associated with it.

\begin{defn}[Compound Element]
  A Compound Element is a representation of a software contract that has been
  defined given a number of other software contracts. Lets assume a software
  contract with the unique name $k$, using the combinator function $f$, with the
  parameters $k_1, k_2, ..., k_n$, that are atomic or compound elements
  identified by their unique name. In such a scenario a Compound Element is
  represented by $k_{f}(k_1, k_2, ..., k_n)$.
\end{defn}

A good example for the compound element is a representation of the logical `and'
when combining two or more contracts. For example suppose we need to express
that a date is valid if it is in an ISO format and occurs after the year 2015.
We can express this as $\mathit{validDate}_{\mathit{and}}(\mathit{validISOdate},
\mathit{after2015})$ assuming the existence of two contracts such as
$\mathit{validISOdate}_{\mathit{atom}}$ and $\mathit{after2015}_{\mathit{atom}}$
for the parameters.

\begin{defn}[Root Element]
  A Root Element is a representation of the software contract that is the
  executable form of the whole regulatory document. It abstracts the executable
  contract representing the whole regulation. \end{defn}

\begin{table}
  \footnotesize
  \begin{center}
    \begin{tabular}{| l | p{7cm} |}
      \hline
      Element & CNL   \\
      \hline
      $k_{atom}$ & The contract $k$ must hold. \\
      $k_{or}(k_1, k_2, ..., k_n)$ & The contract $k$ holds, if at least one of the contracts $k_1, k_2, ..., k_n$ holds. \\
      $k_{and}(k_1, k_2, ..., k_n)$ & The contract $k$ holds, if all of the contracts $k_1, k_2, ..., k_n$ hold. \\
      $k_{keys}(k_1, k_2, ..., k_n)$  & The contract $k$ holds, if for the keys and values of this map the contracts $k_1, k_2, ..., k_n$ hold. \\
      $k_{coll-of}(k_1)$ & The contract $k$ holds, if for the members of this collection the contract $k_1$ holds. \\
      \hline
    \end{tabular}
  \end{center}
  \caption{The elements mapped to the CNL sentences.}

  \label{tab:cnl2text}
\end{table}

\begin{table}
  \footnotesize
  \begin{center}
    \begin{tabular}{| l | p{5.5cm} | |}
      \hline
      Element & Abstracted `clojure.spec' Contract  \\
      \hline
      $k_{atom}$ & (s/def $k$ ...) \\
      $k_{or}(k_1, k_2, ..., k_n)$ & (s/def $k$ (s/or $k_1$ $k_1$ $k_2$ $k_2$ ... $k_n$ $k_n$)) \\
      $k_{and}(k_1, k_2, ..., k_n)$ & (s/def $k$ (s/and $k_1$ $k_2$ ... $k_n$)) \\
      $k_{keys}(k_1, k_2, ..., k_n)$ & (s/def $k$ (s/keys :req [$k_1$ $k_2$ ... $k_n$] )) \\
      $k_{coll-of}(k_1)$ & (s/def $k$ (s/coll-of $k_1$)) \\
      \hline
    \end{tabular}
  \end{center}
  \caption{Showing how the elements map to `clojure.spec' based contracts. It is
    assumed that `s' is the `clojure.spec' namespace while $k, k_1, ..., k_n$
    correspond to namespaced keywords. }
  \label{tab:cnl2spec}
\end{table}

How these semantics are mapped to the CNL and the actual contracts is dependent
on the combinator functions we allow. In particular we use a set of four
functions, that are abstracted versions of the combinator functions found in the
`clojure.spec' library (see Section~\ref{sec:softwarecontracts}). In particular
we use the logical combinators `and' and `or', as well as `keys' and `coll-of'
denoting that the restrictions hold on maps and collections respectively. Note
that these are not quite the same as the spec combinators in the `clojure.spec'
library. One difference is that we abstract away some semantics of the actual
spec combinators. For example in the CNL we have no ability to specify the
length of a collection or differentiate between the required and the optional
keys. The other difference is that certain combinator functions in the spec
library, notably `s/or', offer ways to describe a tag for each parameter. In our
case this will always be the name of the registered spec, hence we can safely
abstract this away.

Table~\ref{tab:cnl2text} shows how the elements are mapped to the actual CNL
sentences, while Table~\ref{tab:cnl2spec} shows how the elements map to
`clojure.spec' contracts. Due to space limitations, as well as the fact that
specs can have additional options (e.g.: optional keys)
beyond what is required to match the semantics of the CNL, the spec descriptions are
abstracted to minimal contracts.

A minimum requirement to demonstrate the viability of this approach hinges on
the question whether it can effectively encode and use actual compliance
regulations.

In the next section, we aim to answer this question, with a particular use case.

\section{Use Case: Money Market Statistical Reporting}\label{sec:usecase}

In order to demonstrate our approach, as described in
Section~\ref{sec:approach}, we aim to answer the question: ``Can we effectively
encode and use the contract system: `clojure.spec', to automate regulations?''
To demonstrate this, we have implemented a portion of the MMSR regulations
\cite{eu-1333-2014}. In particular we implement a part of the regulations
applicable to reporting on the secured market segment, as outlined in Sections
3.3 and 3.4 of version 2.3.1 of the reporting regulations document
\cite{mmsr-reporting}. Such a message consists of a number of fields, where the
regulations specify a set of fields together with descriptions on their accepted
values.

\begin{figure}
  \begin{center}
    \includegraphics[width=0.6\textwidth]{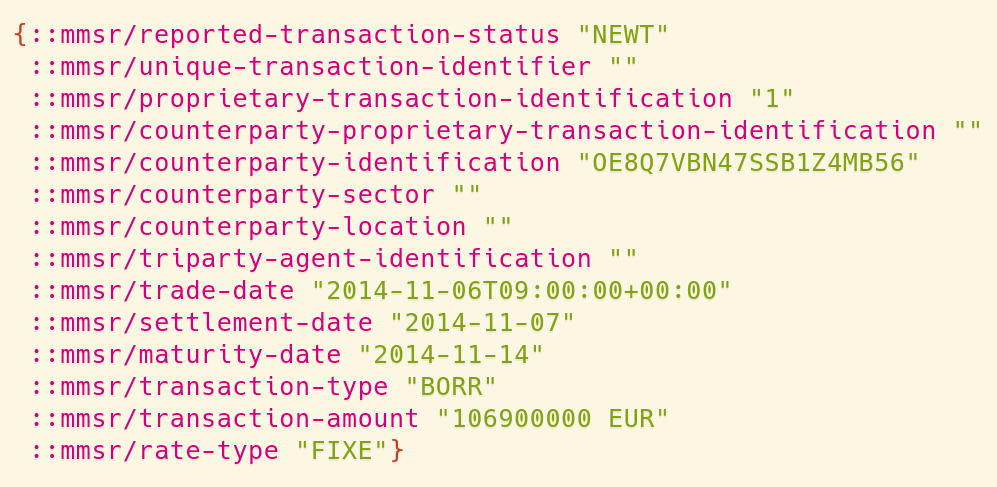}
  \end{center}
  \caption{A portion of the example MMSR message represented as a Clojure map}
  \label{fig:example}
\end{figure}

The data of such a report can be expressed as a map of values in Clojure with
the various attributes as namespaced keywords. Such a representation, which is
based on a portion of an example given in the Annex VII of the MMSR regulations,
can be seen in Figure~\ref{fig:example}.

The full set of conditions on this report is too numerous to describe in detail.
Here we use just a small sample of such requirements on the trade date element
to illustrate our approach.

The relevant text for ensuring that the trade date element is compliant is given by the following extract from MMSR
Reporting Instructions version 2.3.1 page 27 \cite{mmsr-reporting}:

\texttt{``Date time is always represented in an ISO 8601 format in one of three
  forms: YYYY-MM-DDThh:mm:ss+/-hh:mm or \\ YYYY-MM-DDThh:mm:ss.sss+/-hh:mm or
  YYYY-MM-DD. ''}
 
The domain expert who is using this description can transcribe it into a CNL
version of the contract which is as follows:

\begin{itemize}
\item The contract $\textrm{::mmsr/valid-date-time-ms}$ must hold.
\item The contract $\textrm{::mmsr/valid-date-time-no-ms}$ must hold.
\item The contract $\textrm{::mmsr/valid-date}$ must hold.
\item The contract $\textrm{::mmsr/trade-date}$ holds, if at least one of the
  contracts \linebreak $\textrm{::mmsr/valid-date-time-ms},
  \textrm{::mmsr/valid-date-time-no-ms}, \textrm{::mmsr/valid-date}$ holds.
\end{itemize}

This CNL is an abstraction of the actual contracts. Here we show the \\
::mmsr/trade-date contract. As always `s' stands for the `clojure.spec'
namespace, and assuming this is defined in the namespace for mmsr (hence we can
use ``::'' as opposed to ``::mmsr'').

\begin{lstlisting}
                   (s/def ::trade-date (s/or
                    ::valid-date-time-ms ::valid-date-time-ms
                    ::valid-date-time-no-ms ::valid-date-time-no-ms
                    ::valid-date ::valid-date))
\end{lstlisting}

Note that that here we abstracted away some of the implementation details of the
contract, such as the attached generators.

In the CNL-based abstraction of the all contracts, there are 23 atomic elements
and 15 compound ones (one of which functions as the root element). With this
contract based implementation, data representing a reported transaction can be
checked for validity. Similarly with the contracts we can also answer why, or
why not, this validity holds (see Section~\ref{sec:softwarecontracts} for more
details).

In addition to validation, there are generators for many of these contracts as
well. These allow for the generation of a valid example that fulfills the
required conditions. For the more simpler contracts, these are automatically
generated, while in other cases they were created manually using functions from
the generator library.

As mentioned previously the question we aim to answer in this evaluation is:
``Can we effectively encode and use the contract system, `clojure.spec', to
automate regulations?''. In the case of the above mentioned regulations,
effectively encoding the regulations has been demonstrated by expressing the
MMSR regulatory constraints with software contracts, with the CNL functioning as
an intermediary abstraction.

The other part of the question, whether we can effectively use the contract
system, can be answered by demonstrating that it can validate a non-trivial
amount of messages in a reasonable amount of time. To provide some level of
comparison, though the exact checks and conditions might differ, a number of
data quality checks are provided as a service by the European Central Bank
(ECB), that aim to process at a rate of 100 transactions per 10 seconds
\cite{mmsrfaq}.

In addition, as we are aiming to use the generation functionality as part of
interaction with experts, it is important that this aspect of the approach is
measurable and efficient. To this end, we evaluate three functionalities of our
approach: the validation of a particular message, the explanation of why a
particular example is valid and the generation of messages.

In total we measure the execution time of functions in five scenarios:
 
\begin{itemize}
\item Validation: How fast can the example given in this section (in Figure~\ref{fig:example}) be checked for
  validity (true or false)?
\item Conform: How fast can the example given in this section be checked and
  given the explanation on why it is valid?
\item Generation: How fast can a valid message be generated?
\item Generation+Validation: How fast can a message be generated and validated?
\item Generation+Conformance: How fast can a message be generated and explained
  why it is valid?
\end{itemize}

Benchmarking a JVM based language, such as Clojure, has some known pitfalls
\cite{boyer2008robust}, due to the Just-In-Time (JIT) compilation among other
things. To benchmark in a way to mitigate these issues we use the Criterium
\cite{criterium} library. These benchmarks were run on a system running Ubuntu
18.04.1 LTS with an Intel Core i7-8700K and 16 GB of RAM. The benchmarks were
ran using Clojure 1.9, using Java version 1.8.0_222 and the OpenJDK server. In
Criterium the default settings and parameters were used, with the estimated
overhead parameter set at 6.000000 ns.

In Table~\ref{tab:results} we show the results of our experiments, in particular
the mean execution time and the standard deviation. Validating and conforming
seem effective, even on generated cases, and could be used to process a large
volume of messages rapidly. The generation process itself is somewhat more
expensive, with a higher variance due to, we believe, more complex generators
along certain paths, but still effective enough to be able to integrated into a
responsive UI.

\begin{table}
  \begin{center}
    \footnotesize
    \begin{tabular}{ | l | l | l |}
      \hline
      Scenario & Exec. Mean & Std-Deviation \\
      \hline
      Validation & 17.062645 $\mu$s & 331.960205 ns \\
      Conform & 17.137722 $\mu$s &  634.434731 ns \\
      Generation & 469.627774 $\mu$s & 21.070658 $\mu$s \\
      Generation+Validation & 504.692289 $\mu$s & 10.238499 $\mu$s \\
      Generation+Conformance & 505.615074 $\mu$s & 14.397948 $\mu$s \\
      \hline
    \end{tabular}
  \end{center}
  \caption{Results for the scenarios.}
  \label{tab:results}
\end{table}

\section{Related Work \& Discussion}\label{sec:related}

With the rapid increase of regulatory compliance related costs, it is an active
challenge in both research and development to explore a suitable chain of
methodologies and technologies to fully automate the compliance checking process
\cite{hutchison_emerging_2010}. Compliance checking is mainly separated into two
approaches: 1) compliance by design (before-the-fact) and 2) retrospective
reporting (after-the-fact) \cite{alonso_modeling_2007}. Our approach is firmly
in the first category: we hope to facilitate the creation of software that is
compliant by design.

There is a large volume of existing approaches for compliance. Logic-based
techniques encode regulations into a logical formalism for reasoning. They
generally make a trade-off for more limited expressiveness to enable features
such as model consistency checking, as in the work of Kerrigan \& Law
\cite{kerrigan_logic-based_2003} and Awad, Decker \& Weske \cite{dumas_efficient_2008}.
Our approach currently has no such features as the software contracts themselves
are often too expressive for this purpose, while our current CNL has a
very limited set of operators and structures. Nonetheless if we expand the CNL with more
operators, we can provide such model checking features on the CNL in the future.

Rule-based approaches have been extensively used for compliance checking, for
example in the work of Strano, Molina-Jimenez \& Shrivastava \cite{strano2009implementing}. They can provide good
expressiveness for tackling compliance. Our approach offers a stronger
integration with the software development methodologies, as contract
systems are more integrated within a number of programming languages than rule engines.
In addition the CNL provides an abstraction layer for domain experts that could
otherwise be overwhelmed when dealing with an expressive rule system.
However, rule engines can provide considerably more control over the execution
of the regulations
and the interpretation of facts than a system such as `clojure.spec'. For
example certain cases could result in a `warning' or `error' as opposed to just
a boolean `valid' or `invalid' indicator. If such features would be added to the
CNL, mapping this artifact to rule engines could be an additional step in our approach.

Finally semantic technologies are another inspiration for compliance checking~\cite{zhong_ontology-based_2012,pieter_semantic_2015}. The main benefit is that
they provide facilities to use, reuse and publish knowledge. Unfortunately
the use of Semantic Web technologies in software development, in
practical contexts, is currently limited. However, by adding the ability to
express concepts in the CNL in the future, we can integrate the advantages of semantic
technologies within our approach.

A key aspect for discussion within our approach is what structures we allow for
the CNL and what do we abstract away from the contract implementation. On one hand our CNL is very
limited, especially with regards to the logical operators, and even in the
features for handling data. On the other hand we have found from informal evaluations with domain
experts that these logical operators alone can greatly help in breaking down the
structure of the document and provide insight to the implementation. In addition, handling the
data-flow, even in such limited way, can often provide challenges in
understanding for non-software engineers. As a future
work, more formal evaluation is
required to measure the impact of certain structures in the CNL from the
perspective of the domain experts.

\section{Conclusion and Future Works}\label{sec:conclusion}

In this work we aimed to tackle two major issues with automated approaches to
financial compliance regulations: the ability to express arbitrary regulations
and the issues with involving the software domain as an integral and continuous
stakeholder in this process. To solve these issues we presented an approach
inspired by software contracts, namely the `clojure.spec' library. Our approach
contributes in three directions: it provides strong expressiveness, reasonable performance,
and a systematic methodology to integrate compliance-aware design with the
software development process. In particular, our approach can facilitate
features such as compliance validation with explanations and data
generation. These features can provide domain experts more insights into the
automation of regulations, as well as artifacts to ensure compliance within software systems.

In the future we aim to apply this approach to more regulations.
In addition, we hope to explore and expand the formalism behind the
CNL to further enable domain experts within the compliance
automation process. 

\bibliographystyle{splncs04}
\bibliography{compliance}

\begin{thebibliography}{10}
\providecommand{\url}[1]{\texttt{#1}}
\providecommand{\urlprefix}{URL }
\providecommand{\doi}[1]{https://doi.org/#1}

\bibitem{criterium}
Criterium. \url{https://github.com/hugoduncan/criterium}, accessed: 2018-07-19

\bibitem{plumatic-schema}
{Plumatic Schema}. \url{https://github.com/plumatic/schema}, accessed:
  2018-07-19

\bibitem{alexandrescu2010d}
Alexandrescu, A.: {The D Programming Language}. Addison-Wesley Professional
  (2010)

\bibitem{dumas_efficient_2008}
Awad, A., Decker, G., Weske, M.: {Efficient Compliance Checking Using BPMN-Q
  and Temporal Logic}. In: Dumas, M., Reichert, M., Shan, M.C. (eds.) Business
  Process Management, pp. 326--341. Springer Berlin Heidelberg (2008)

\bibitem{becker2012generalizability}
Becker, J., Delfmann, P., Eggert, M., Schwittay, S.: {Generalizability and
  Applicability of Model-Based Business Process Compliance-Checking Approaches
  --- A State-of-the-Art Analysis and Research Roadmap}. Business Research
  \textbf{5}(2),  221--247 (Nov 2012)

\bibitem{boyer2008robust}
Boyer, B.: {Robust Java Benchmarking}. IBM Developer Works  (2008)

\bibitem{eu-1333-2014}
{Council of European Union}: {Council Regulation ({EU}) 2014/1333} (2014)

\bibitem{eu-679-2016}
{Council of European Union}: {Council Regulation ({EU}) 2016/679} (2016)

\bibitem{mmsr-reporting}
{Council of European Union}: {Reporting Instructions for the Electronic
  Transmission of Money Market Statistical Reporting (MMSR)} (2017), version
  2.3.1

\bibitem{world2014rdf}
Cyganiak, R., Wood, D., Lanthaler, M., Klyne, G., Carroll, J.J., McBride, B.:
  {RDF 1.1 Concepts and Abstract Syntax}. W3C Recommendation  (2014)

\bibitem{mmsrfaq}
{European Central Bank}: {Money Market Statistical Reporting - Questions and
  Answers Version 3.1}

\bibitem{fahndrich2010static}
F{\"a}hndrich, M.: {Static Verification for Code Contracts}. In: Cousot, R.,
  Martel, M. (eds.) Static Analysis. pp.~2--5. Springer Berlin Heidelberg
  (2010)

\bibitem{fang2017approach}
Fang, M., Tompits, H.: {An Approach for Representing Answer Sets in Natural
  Language}. In: Seipel, D., Hanus, M., Abreu, S. (eds.) Declarative
  Programming and Knowledge Management, pp. 115--131. Springer International
  Publishing (2018)

\bibitem{felleisen2015racket}
Felleisen, M., Findler, R.B., Flatt, M., Krishnamurthi, S., Barzilay, E.,
  McCarthy, J., Tobin-Hochstadt, S.: {The Racket Manifesto}. In: LIPIcs-Leibniz
  International Proceedings in Informatics. vol.~32. Schloss
  Dagstuhl-Leibniz-Zentrum fuer Informatik (2015)

\bibitem{fitzgerald2014continuous}
Fitzgerald, B., Stol, K.J.: {Continuous Software Engineering and Beyond: Trends
  and Challenges}. In: Proceedings of the 1st International Workshop on Rapid
  Continuous Software Engineering. pp.~1--9. RCoSE 2014, ACM, New York, NY, USA
  (2014)

\bibitem{attempto1}
Fuchs, N.E., Kaljurand, K., Kuhn, T.: {Attempto Controlled English for
  Knowledge Representation}. In: Reasoning Web, pp. 104--124. Springer (2008)

\bibitem{grasshoff2017global}
Grasshoff, G., Mogul, Z., Pfuhler, T., Gittfried, N., Wiegand, C., Bohn, A.,
  Vonhoff, V.: {Global Risk 2017: Staying the Course in Banking}. Boston
  Consulting Group  (2017)

\bibitem{hickey2008clojure}
Hickey, R.: {The Clojure Programming Language}. In: Proceedings of the 2008
  Symposium on Dynamic languages. p.~1. ACM (2008)

\bibitem{clojure-spec}
Hickey, R.: {clojure.spec - Rationale and Overview}.
  \url{https://clojure.org/about/spec} (2017), accessed: 2018-07-30

\bibitem{hoare1969axiomatic}
Hoare, C.A.R.: {An Axiomatic Basis for Computer Programming}. Communications of
  the ACM  \textbf{12}(10),  576--580 (1969)

\bibitem{jazequel1997design}
Jazequel, J.M., Meyer, B.: {Design by Contract: The Lessons of Ariane}.
  Computer  \textbf{30}(1),  129--130 (1997)

\bibitem{kerrigan_logic-based_2003}
Kerrigan, S., Law, K.H.: {Logic-Based Regulation Compliance-Assistance}.
  p.~126. ACM Press (2003)

\bibitem{leavens2006preliminary}
Leavens, G.T., Baker, A.L., Ruby, C.: {Preliminary Design of JML: A Behavioral
  Interface Specification Language for Java}. ACM SIGSOFT Software Engineering
  Notes  \textbf{31}(3),  1--38 (2006)

\bibitem{meyer1992applying}
Meyer, B.: {Applying 'Design by Contract'}. Computer  \textbf{25}(10),  40--51
  (1992)

\bibitem{meyer1992eiffel}
Meyer, B.: {Eiffel: The Language}. Prentice Hall (1992)

\bibitem{clojure-spec-guide}
Miller, A.: {Spec Guide}. \url{https://clojure.org/guides/spec}, accessed:
  2018-07-19

\bibitem{pieter_semantic_2015}
Pieter, P., Zhang, S.: Semantic {Rule}-checking for {Regulation} {Compliance}
  {Checking}: {An} {Overview} of {Strategies} and {Approaches}. In: 32rd
  international {CIB} {W}78 conference, {Proceedings}. Eindhoven, The
  Netherlands (2015)

\bibitem{alonso_modeling_2007}
Sadiq, S., Governatori, G., Namiri, K.: Modeling {Control} {Objectives} for
  {Business} {Process} {Compliance}. In: Business {Process} {Management},
  vol.~4714, pp. 149--164. Springer Berlin Heidelberg (2007)

\bibitem{strano2009implementing}
Strano, M., Molina-Jimenez, C., Shrivastava, S.: {Implementing a Rule-Based
  Contract Compliance Checker}. In: Conference on e-Business, e-Services and
  e-Society. pp. 96--111. Springer (2009)

\bibitem{hutchison_emerging_2010}
Syed~Abdullah, N., Sadiq, S., Indulska, M.: Emerging {Challenges} in
  {Information} {Systems} {Research} for {Regulatory} {Compliance}
  {Management}. In: Advanced {Information} {Systems} {Engineering}, vol.~6051,
  pp. 251--265. Springer Berlin Heidelberg (2010)

\bibitem{zhong_ontology-based_2012}
Zhong, B., Ding, L., Luo, H., Zhou, Y., Hu, Y., Hu, H.: {Ontology-Based
  Semantic Modeling of Regulation Constraint for Automated Construction Quality
  Compliance Checking}. Automation in Construction  \textbf{28},  58--70 (Dec
  2012)

\end{thebibliography}
 
\end{document}